\documentclass[prd,12pt,tightenlines,preprintnumbers,showpacs,%
superscriptaddress,nofootinbib,amsmath,amssymb]{revtex4}
\usepackage{bm}
\usepackage{graphicx}

\begin{document}

\preprint{CECS-PHY-04/04}
\preprint{hep-th/0403228}

\title{Stealth Scalar Field Overflying a $2+1$ Black Hole}

\author{Eloy Ay\'on--Beato}\email{ayon-at-cecs.cl}
\affiliation{Centro~de~Estudios~Cient\'{\i}ficos~(CECS),
~Casilla~1469,~Valdivia,~Chile.}
\affiliation{Departamento~de~F\'{\i}sica,~CINVESTAV--IPN,%
~Apdo.~Postal~14--740,~07000,~M\'exico~D.F.,~M\'exico.}
\author{Cristi\'{a}n Mart\'{\i}nez}\email{martinez-at-cecs.cl}
\author{Jorge Zanelli}\email{jz-at-cecs.cl}
\affiliation{Centro~de~Estudios~Cient\'{\i}ficos~(CECS),
~Casilla~1469,~Valdivia,~Chile.}

\begin{abstract}
A nontrivial scalar field configuration of vanishing
energy--momentum is reported. These matter configurations have no
influence on the metric and therefore they are not be ``detected"
gravitationally. This phenomenon occurs for a time--dependent
nonminimally coupled and self--interacting scalar field on the
$2+1$ (BTZ) black hole geometry. We conclude that such stealth
configurations exist for the static $2+1$ black hole for any value
of the nonminimal coupling parameter $\zeta\neq0$ with a fixed
self--interaction potential $U_\zeta(\Phi)$. For the range
$0<\zeta\leq1/2$ potentials are bounded from below and for the
range $0<\zeta<1/4$ the stealth field falls into the black hole
and is swallowed by it at an exponential rate, without any
consequence for the black hole.
\end{abstract}


\maketitle

One of the hallmarks of General Relativity is the fact that the
presence of matter can be detected by the metric. For instance,
the total mass--energy contained in any form of matter that falls
into a black hole can be revealed by the metric at infinity
through the famous ADM formula. More generally, the role of matter
in gravity is to provide a source for the curvature of spacetime
through the energy--momentum tensor. This general feature can have
exceptions, as we report here, if the nontrivial field content and
the background geometry are so special that the energy--momentum
tensor vanishes identically and therefore, the spacetime geometry
can be exactly the same as the one that solves the matter--free
Einstein equations.

Alberto Garc\'{\i}a, for one, has made it a profession of dressing
spacetime geometries with all sorts of exotic and nonstandard
drapery. It is only fitting therefore, to contribute to a volume
in his honor this stealth form of matter, which would certainly
not escape his detection.

\centerline{------}

Consider a self--interacting scalar field nonminimally coupled to
$2+1$ gravity described by the action \cite{Ayon-Beato:2001sb}
\begin{equation}\label{eq:ac}
S=\int{d^3x}\sqrt{-g} \left(\frac1{2\kappa}\left(R+2l^{-2}\right)
-\frac12\nabla_\mu\Psi\nabla^\mu\Psi -\frac12\zeta\,R\,\Psi^2
-U(\Psi)\right),
\end{equation}
where $\Lambda=-l^{-2}$ is the cosmological constant, $\zeta$ is
the nonminimal coupling parameter and $U(\Psi)$ is the
self--interaction potential. The corresponding field equations are
\begin{equation}\label{eq:Ein}
G_\mu^{~\nu}-l^{-2}\delta_\mu^{~\nu}={\kappa}T_\mu^{~\nu},
\end{equation}
and
\begin{equation}\label{eq:scalar}
\Box\Psi=\zeta\,R\,\Psi +\frac{\mathrm{d}U(\Psi)}{\mathrm{d}\Psi},
\end{equation}
where the energy--momentum tensor is given by
\begin{equation}\label{eq:T}
T_\mu^{~\nu}=\nabla_\mu\Psi\nabla^\nu\Psi
-\delta_\mu^{~\nu}\left(\frac12\nabla_\alpha\Psi\nabla^\alpha\Psi+U(\Psi)\right)
+\zeta\left(\delta_\mu^{~\nu}\Box-\nabla_\mu\nabla^\nu
+G_\mu^{~\nu}\right)\Psi^2 .
\end{equation}

We are interested in gravitationally undetectable configurations
in the sense defined in the first paragraph, i.e., nontrivial
solutions to the field equations (\ref{eq:Ein}) and
(\ref{eq:scalar}) such that $T_\mu^{~\nu}=0$. For these
configurations both sides of Einstein equations vanish
independently. The only symmetry we shall impose is that the
solutions possess cyclic symmetry, that is, they must be invariant
under the action of the 1--parameter group $SO(2)$. Under these
conditions the metric must be a cyclic solution of the $2+1$
vacuum Einstein equations with a negative cosmological constant,
\[
G_\mu^{~\nu}-l^{-2}\delta_\mu^{~\nu}=0.
\]
By Birkhoff's theorem in $2+1$ dimensions \cite{Ayon-Beato:2004if}
the geometry must be given by the so--called BTZ black hole solution
\cite{Banados:wn,Banados:1992gq}\footnote{We shall not consider here
the self--dual Coussaert--Henneaux spacetimes
\cite{Coussaert:1994tu}, which are also allowed by Birkhoff's
theorem (see \cite{Ayon-Beato:2004if}), since it can be shown that
in this case there are no stealth configurations for which
$T_\mu^{~\nu}=0$.}
\begin{equation}\label{eq:BTZ}
\bm{g}_{\mathrm{BTZ}}=-F(r)\bm{dt}^2+\frac{\bm{dr}^2}{F(r)}
+r^2\left(\bm{d\phi}-\frac{J}{2r^2}\bm{dt}\right)^2,
\end{equation}
with
\begin{equation}\label{eq:F(r)}
F(r)\equiv\frac{r^2}{l^2}-M+\frac{J^2}{4r^2},
\end{equation}
where $M$ and $J$ are the mass and angular momentum of the black
hole, respectively, with $|J|\leq{Ml}$. In these coordinates the
cyclic symmetry is generated by the Killing field
$\bm{m}=\bm{\partial_\phi}$ and the cyclic invariance of the
scalar field is expressed in the dependence $\Psi=\Psi(t,r)$.

In the present context a stealth configuration is a nontrivial
solution $\Psi$ such that
$T_\mu^{~\nu}(\bm{g}_{\mathrm{BTZ}},\Psi)=0$. Using definition
(\ref{eq:T}), these equations can be explicitly written for the
metric (\ref{eq:BTZ}) as
\begin{subequations}\label{eq:Tc}
\begin{eqnarray}
\label{eq:Ptt} T_t^{~t}-T_{\phi}^{~\phi}
=\frac{\Psi\partial_{t}\Psi}{F}\left(
2\zeta\frac{\partial^2_{tt}\Psi}{\partial_{t}\Psi}
-(1-2\zeta)\frac{\partial_{t}\Psi}{\Psi}
-2\zeta\frac{{M}F}{r}\frac{\partial_{r}\Psi}{\partial_{t}\Psi}\right)
          &=&0,\\
\nonumber & &  \\
\label{eq:Ptr} T_t^{~r}=-F\Psi\partial_{t}\Psi\left(
2\zeta\frac{\partial^2_{rt}\Psi}{\partial_{t}\Psi}
-(1-2\zeta)\frac{\partial_{r}\Psi}{\Psi}
-2\zeta\frac{r}{l^2F}\right)
          &=&0,\\
\nonumber & &  \\
\label{eq:Pr} T_\phi^{~~\!t}=J\zeta\frac{\Psi\partial_{r}\Psi}{r}
          &=&0, \\
\nonumber & &  \\
\label{eq:Pt} T_\phi^{~~\!r}=-J\zeta\frac{\Psi\partial_{t}\Psi}{r}
          &=&0, \\
\nonumber & &  \\
\label{eq:Prr}
T_{\phi}^{~\phi}-T_r^{~r}=F\Psi\partial_{r}\Psi\left(
2\zeta\frac{\partial^2_{rr}\Psi}{\partial_{r}\Psi}
-(1-2\zeta)\frac{\partial_{r}\Psi}{\Psi}
+\frac{\zeta(4Mr^2-J^2)}{2r^3F}\right)
          &=&0,
\end{eqnarray}
\begin{eqnarray}
\label{eq:U} T_\phi^{~\phi}-T_r^{~r}-T_t^{~t}=
U(\Psi)-\frac{\Psi^2}{2}\left(F\frac{(\partial_{r}\Psi)^2}{\Psi^2}
+\frac{8\zeta(r^2-Ml^2)}{l^2r}\frac{\partial_{r}\Psi}{\Psi}
-\frac{1}{F}\frac{(\partial_{t}\Psi)^2}{\Psi^2}
+\frac{2\zeta}{l^2}\right)
          &=&0.\quad\,\,\,\,~
\end{eqnarray}
\end{subequations}

There is no need to solve the scalar equation (\ref{eq:scalar}) as
it is automatically satisfied as a consequence of the conservation
of the energy--momentum tensor (\ref{eq:T}). From
Eqs.~(\ref{eq:Pr}) and (\ref{eq:Pt}), one concludes that the
existence of nontrivial configurations requires that the
background geometry be a non--rotating black hole, $J=0$. (Another
possibility following from Eqs.~(\ref{eq:Pr}) and (\ref{eq:Pt}) is
that the scalar field be minimally coupled, $\zeta=0$, but this
condition would imply the trivial solution
$\Psi(t,r)=\mathrm{const.}$)

For $J=0$ and $\zeta\ne1/4$, Eq.~(\ref{eq:Prr}) can be
straightforwardly integrated, giving
\begin{equation}\label{eq:Psi(f,h)}
\Psi(t,r)=\left(f(t)\sqrt{r^2-Ml^2}+h(t)\right)^{-2\zeta/(1-4\zeta)},
\end{equation}
where $f(t)$ and $h(t)$ are integration functions. This result is
further restricted by Eqs.~(\ref{eq:Ptr}) and (\ref{eq:Ptt})
giving $h(t)=h=\mathrm{const.}$ and
\begin{equation}\label{eq:feq}
\frac{\mathrm{d}^2f}{\mathrm{d}t^2}-\frac{M}{l^2}f=0,
\end{equation}
respectively. Hence, the final expression for the scalar field
reads
\begin{equation}\label{eq:Psi(t,r)}
\Psi(t,r)=\left[K\cosh\left(\frac{\sqrt{M}}{l}(t-t_0)\right)
\sqrt{r^2-Ml^2}+h\right]^{-2\zeta/(1-4\zeta)},
\end{equation}
for $M\neq0$, and
\begin{equation}\label{eq:Psi(t,r)M=0}
\Psi(t,r)=\left[K(t-t_0)r+h\right]^{-2\zeta/(1-4\zeta)},
\end{equation}
for $M=0$, where $K$, $h$, and $t_0$ are now integration
constants. Inserting these expressions in Eq.~(\ref{eq:U}) and
using Eqs.~(\ref{eq:Psi(t,r)}) and (\ref{eq:Psi(t,r)M=0}) once
again, the generic self--interaction potential allowing the
existence of a stealth configuration must be of the form
\begin{equation}\label{eq:U(Psi)}
U_\zeta(\Psi)=\frac{\zeta\Psi^2}{l^2(1-4\zeta)^2}
\left(2\zeta\lambda\left|\Psi\right|^{(1-4\zeta)/{\zeta}}
+4\zeta(1-8\zeta)h\left|\Psi\right|^{(1-4\zeta)/(2\zeta)}
+(1-8\zeta)(1-6\zeta)\right),
\end{equation}
where $\lambda=h^2+K^2Ml^2$ for $M\neq0$ and $\lambda=h^2-K^2l^4$
for $M=0$.

The above potential has three parameters: $\zeta$, $\lambda$, and
$h$. The solutions are characterized by two integration constants
the mass $M$ and $t_0$, which can be eliminated by a time
translation. The constant $K$ is clearly not an independent
integration constant, but it is a function of the coupling
constants appearing in the action and the black hole mass.

The case $\zeta=1/8$ is exceptional because not only the
nonminimal coupling is conformally invariant but the
self--interaction potential reduces to
$8l^2U_{1/8}(\Psi)=\lambda\Psi^6$ which is also conformally
invariant in $2+1$ dimensions. In this case the solutions have an
additional integration constant, $h$, which cannot be related to
the coupling constants appearing in the action. The occurrence of
this new integration constant could be related to the conformal
invariance of the matter sector.

\begin{figure}[h]
  \includegraphics[width=5.4cm]{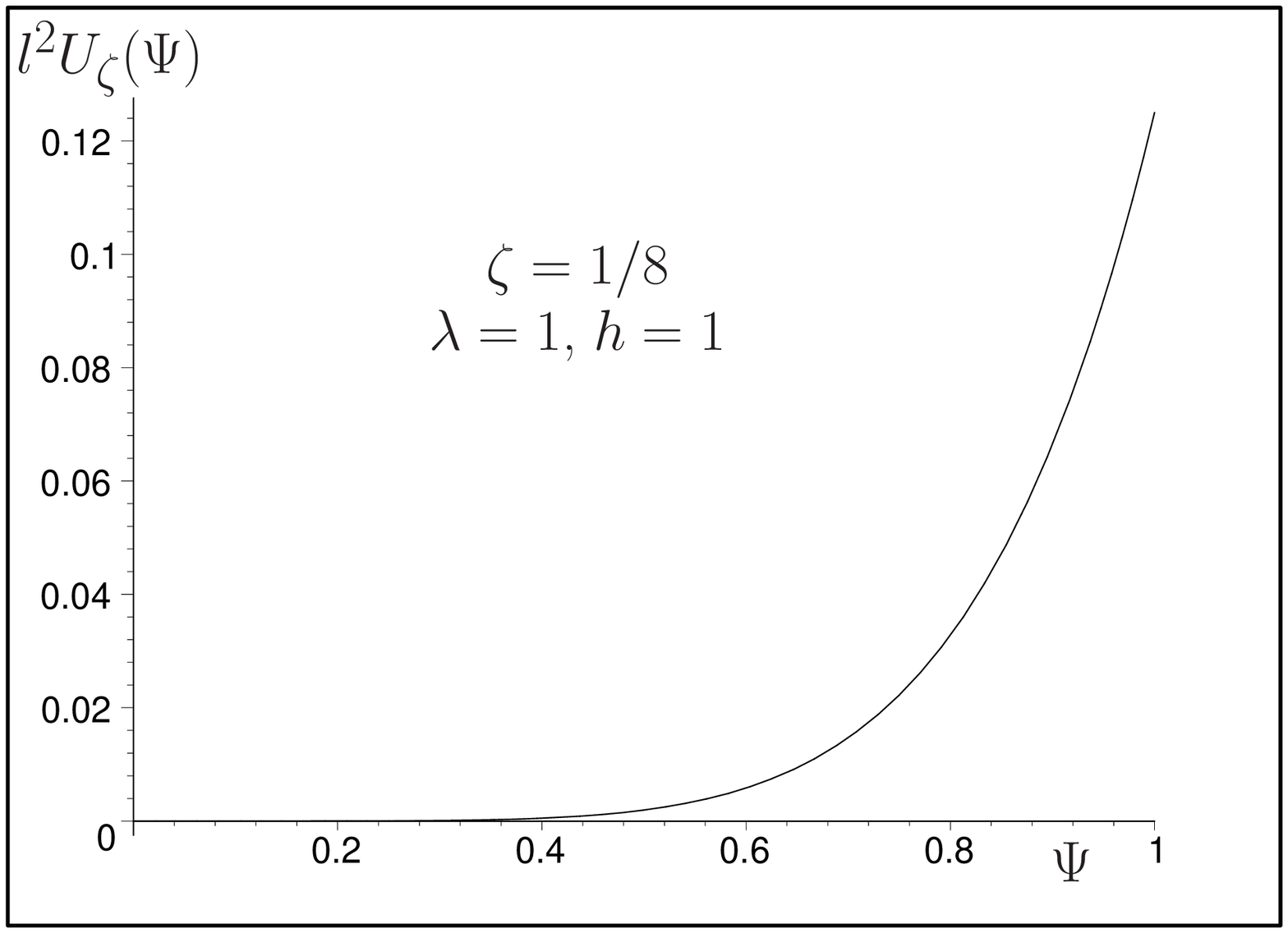}
  \includegraphics[width=5.4cm]{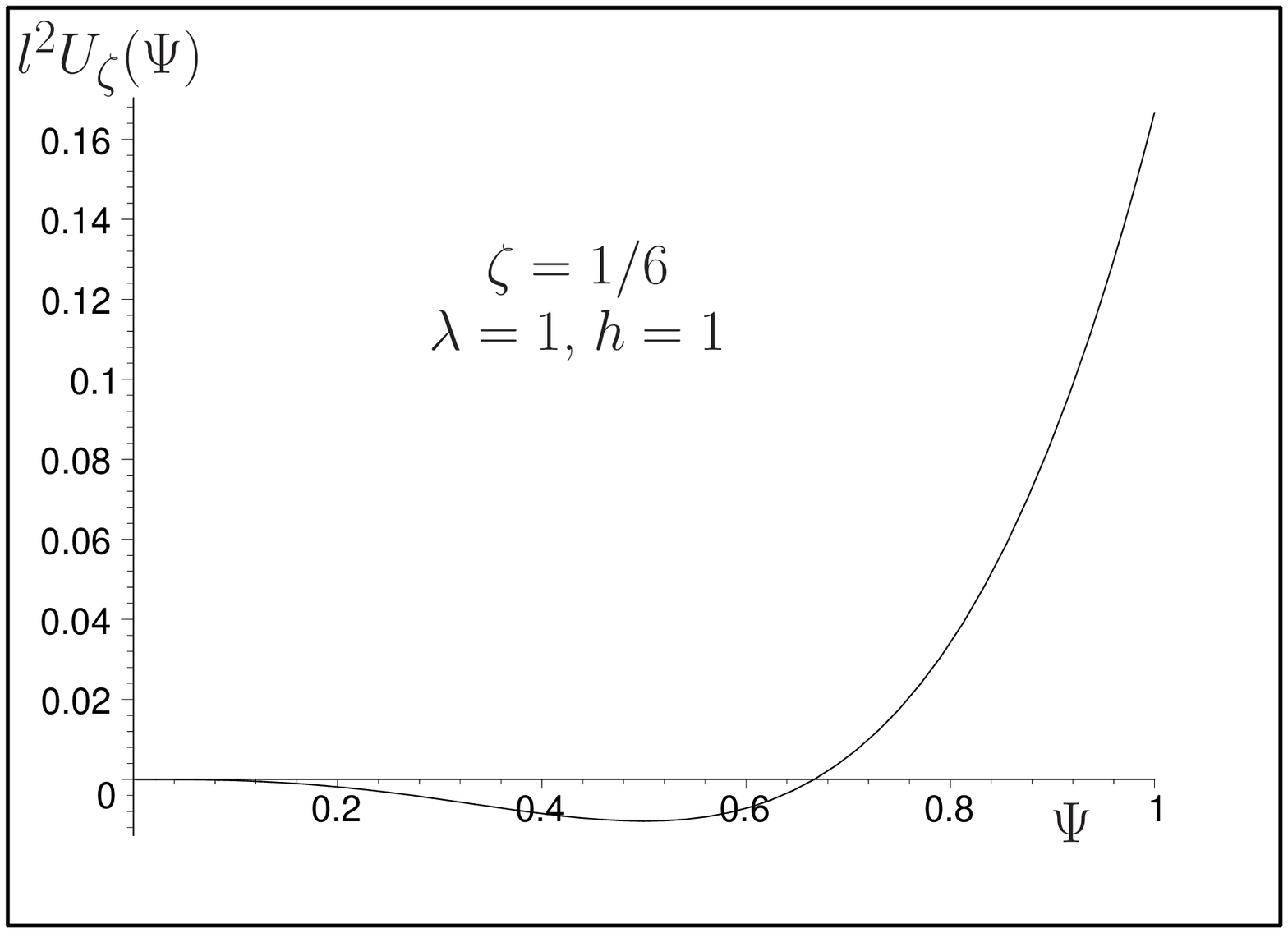}
  \includegraphics[width=5.4cm]{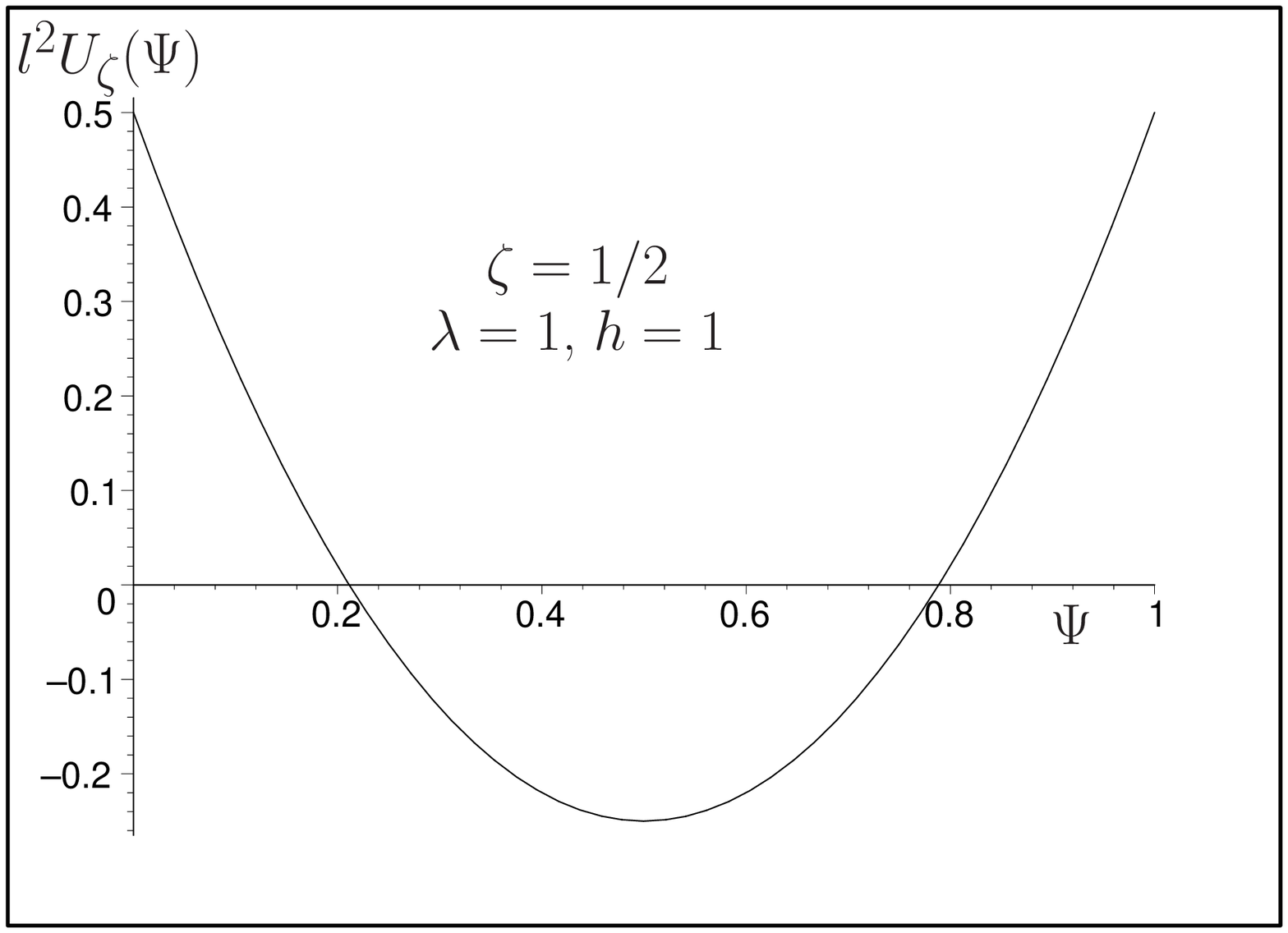}
\caption{\label{fig:Pot}The self--interaction potential
(\ref{eq:U(Psi)}) for conformal coupling ($\zeta=1/8$), and for
the nonminimal couplings $\zeta=1/6$ and $\zeta=1/2$ respectively.
The coupling constants in the potential are fixed as $\lambda=1$
and $h=1$.}
\end{figure}

Examples of the above potentials for different nonminimal coupling
are shown in FIG.~\ref{fig:Pot}.

We have shown that for any nonvanishing value of the nonminimal
coupling parameter $\zeta\neq1/4$, Eqs.~(\ref{eq:Ein}) and
(\ref{eq:scalar}) with self--interaction potential
(\ref{eq:U(Psi)}) have a nontrivial solution given by the static
black hole [(\ref{eq:BTZ}) with $J=0$] for the geometry and by the
time--dependent expression (\ref{eq:Psi(t,r)}) for the scalar
field.

For $\zeta=1/4$ the expression (\ref{eq:Psi(f,h)}) is ill defined,
this is due to the fact that in this case Eq.~(\ref{eq:Prr}) is on
a logarithmic branch, as can be seen from its first integral
\begin{equation}\label{eq:fiPrr}
\frac{\partial_{r}\Psi}{\Psi^{(1-2\zeta)/(2\zeta)}}
=\frac{r\tilde{f}(t)}{\sqrt{r^2-Ml^2}}.
\end{equation}
Clearly, for $\zeta=1/4$ the left hand side integrates as a
logarithm, giving for the scalar field
\begin{equation}\label{eq:Psi(f,h)1/4}
\Psi(t,r) =
\tilde{h}(t)\exp\left(\tilde{f}(t)\sqrt{r^2-Ml^2}\right).
\end{equation}
Using now Eqs.~(\ref{eq:Ptr}) and (\ref{eq:Ptt}) evaluated for
$\zeta=1/4$ we conclude that $\tilde{h}(t)$ is constant and
$\tilde{f}$ satisfies Eq.~(\ref{eq:feq}), as in the generic case.
Hence, for $\zeta=1/4$ the scalar field is
\begin{equation}\label{eq:Psi(t,r)1/4}
\Psi(t,r) =
\Psi_0\exp\left[K\cosh\left(\frac{\sqrt{M}}{l}(t-t_0)\right)
\sqrt{r^2-Ml^2}\right],
\end{equation}
when $M\neq0$, and
\begin{equation}\label{eq:Psi(t,r)1/4M=0}
\Psi(t,r) = \Psi_0\exp\left[K(t-t_0)r\right],
\end{equation}
for $M=0$. These expressions, together with Eq.~(\ref{eq:U}),
imply a non--polynomial form for the self--interaction potential
\begin{equation}\label{eq:U(Psi)1/4}
U_{1/4}(\Psi) = \frac{\Psi^2}{2l^2}\left\{
\left[\ln{\left(\frac{\Psi}{\Psi_0}\right)}+1\right]^2
+\lambda_1-\frac12\right\},
\end{equation}
where $\lambda_1=K^2Ml^2$ for $M\neq0$ and $\lambda_1=-K^2l^4$ for
$M=0$. This potential is shown in FIG.~\ref{fig:Psi(v,r)1/4} for
different values of the constant $\lambda_1$. It should be noticed
that although this potential is bounded from below, the scalar
field behaves explosively (as an exponential for $M=0$ and as the
exponential of an exponential for $M\neq0$) for large times.

\begin{figure}[h]
  \includegraphics[width=7cm]{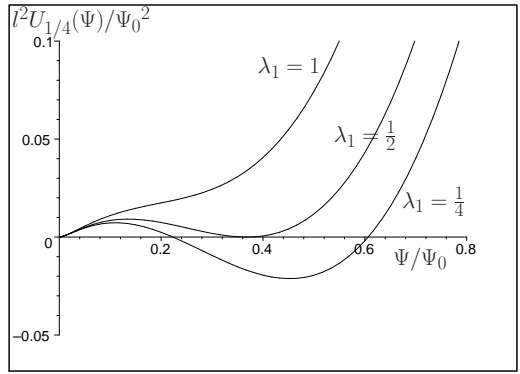}
\caption{\label{fig:Psi(v,r)1/4}Logarithmic self--interaction
potential (\ref{eq:U(Psi)1/4}) for the nonminimal coupling
$\zeta=1/4$. The form of the potential is exhibited for different
values of the constant $\lambda_1$.}
\end{figure}

We are only interested in values of $\zeta$ which give rise to
physically reasonable configurations. The values $\zeta<0$ and
$\zeta>1/2$ should be discarded as they produce self--interaction
potentials which are unbounded from below. On the other hand, for
$1/4\leq\zeta\leq1/2$ the scalar field presents an explosive
growth in time. Hence, we restrict our analysis to the nonminimal
coupling parameters lying in the range $0<\zeta<1/4$.

The physical interpretation of this solution for the black hole
case $M\neq0$ (in the range $0<\zeta<1/4$) can be better
understood from its behavior in ingoing Eddington--Finkelstein
coordinates $(v=t-t_0+r^*,r,\phi)$, which for the static black
hole are given by
\begin{equation}\label{eq:v}
v=t-t_0+r^*
=t-t_0+\frac{l}{2\sqrt{M}}\ln\left(\frac{r-\sqrt{M}l}{r+\sqrt{M}l}\right).
\end{equation}
In these coordinates, the scalar field is expressed as
\begin{equation}\label{eq:Psi(v,r)}
\Psi(v,r)=\left\{K\left[r\cosh\left(\frac{\sqrt{M}}{l}v\right)
+\sqrt{M}l\sinh\left(\frac{\sqrt{M}}{l}v\right)\right]
+h\right\}^{-2\zeta/(1-4\zeta)},
\end{equation}
and in contrast with expression (\ref{eq:Psi(t,r)}), it is evidently
smooth at the horizon $r_+=\sqrt{M}l$ for all times.\footnote{We
thank Viqar Husain for helping us to elucidate this point.} A
graphic description of this expression, in the range $0<\zeta<1/4$,
for different times starting from $v=0$ is shown below in
FIG.~\ref{fig:Psi(v,r)}. From this it can be seen that a smooth
initial scalar field configuration starts to fall into the black
hole and is eventually swallowed by it.
\begin{figure}[h]
  \includegraphics[width=5.4cm]{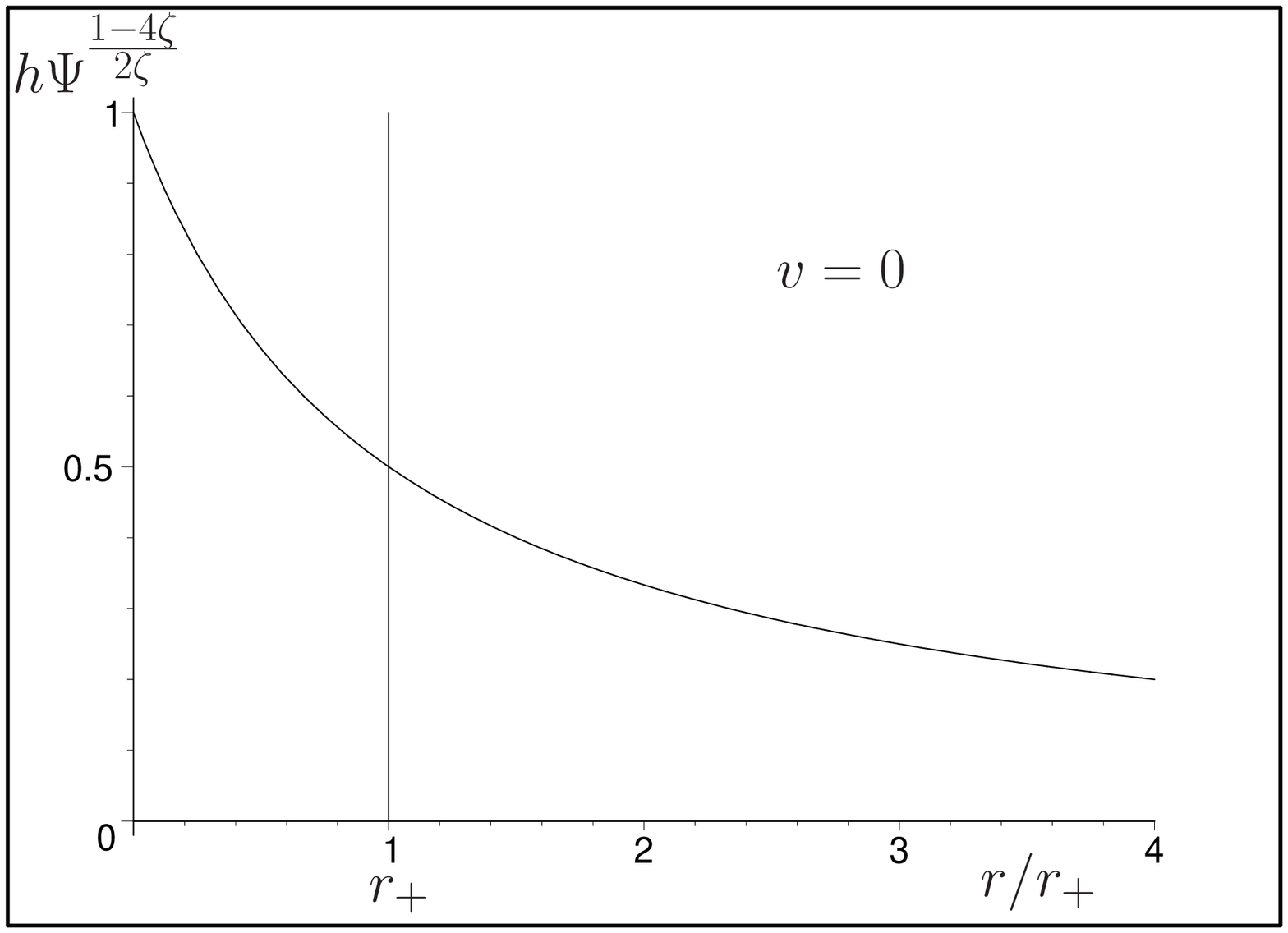}
  \includegraphics[width=5.4cm]{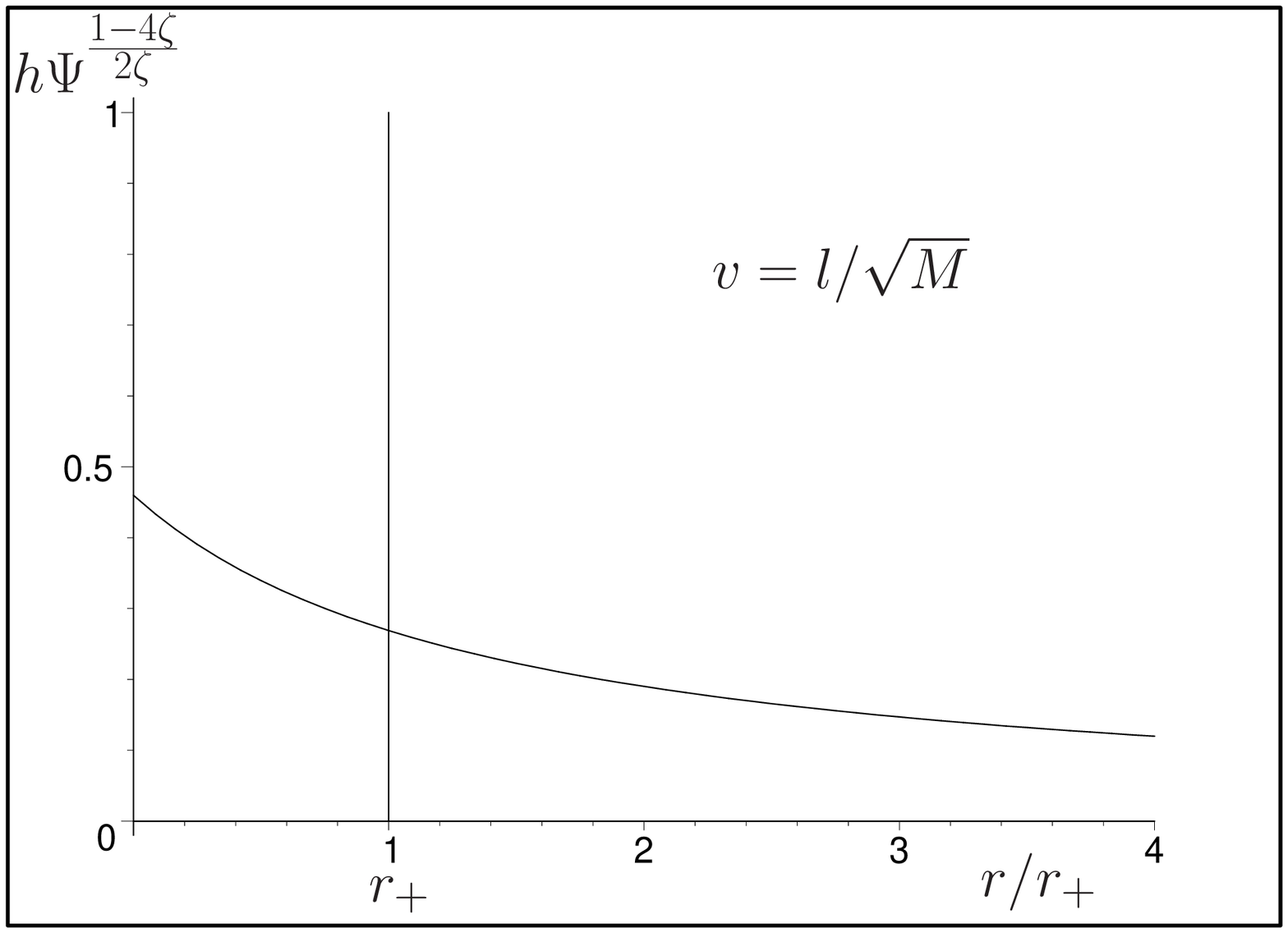}
  \includegraphics[width=5.4cm]{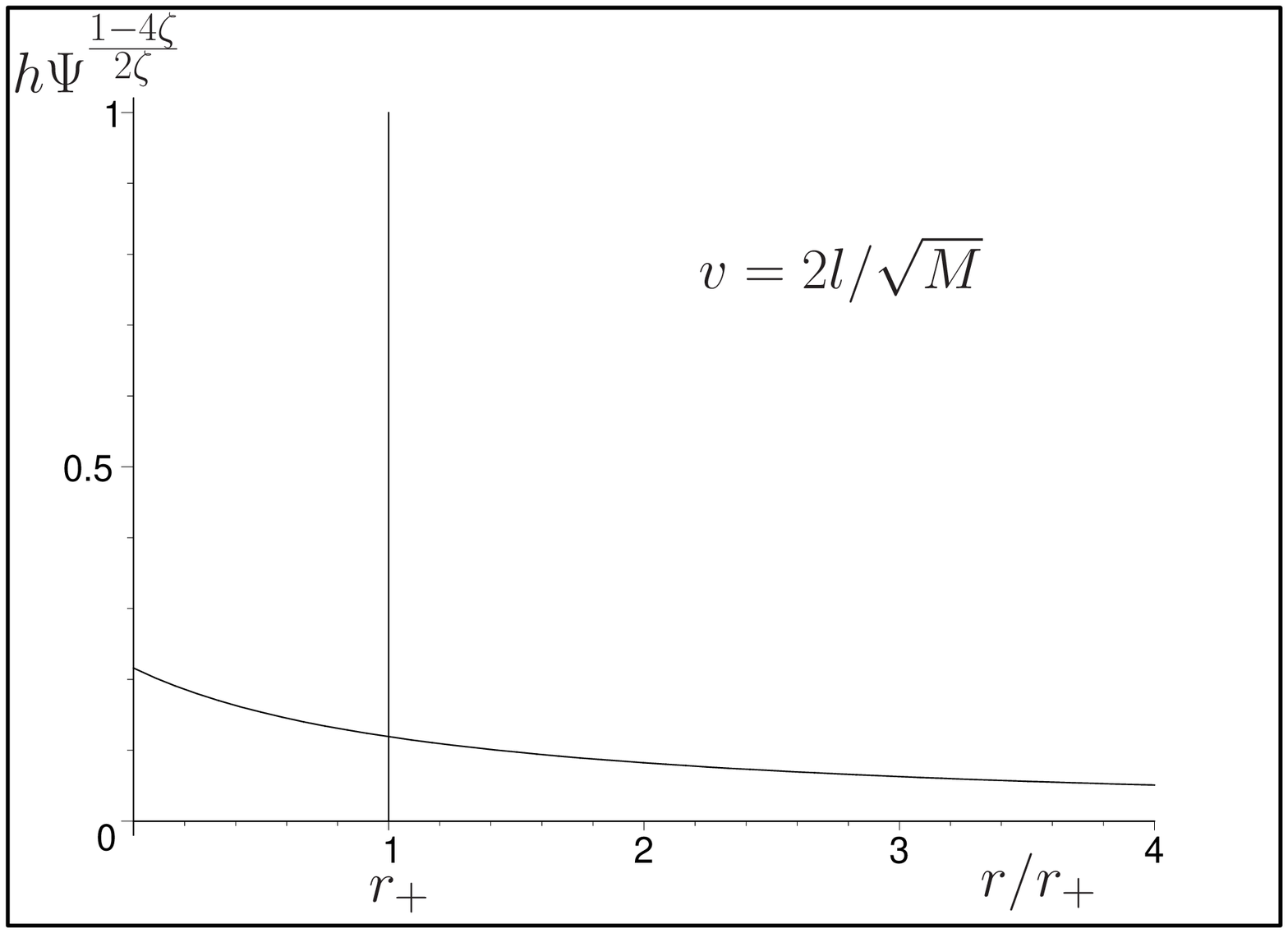}
  \includegraphics[width=5.4cm]{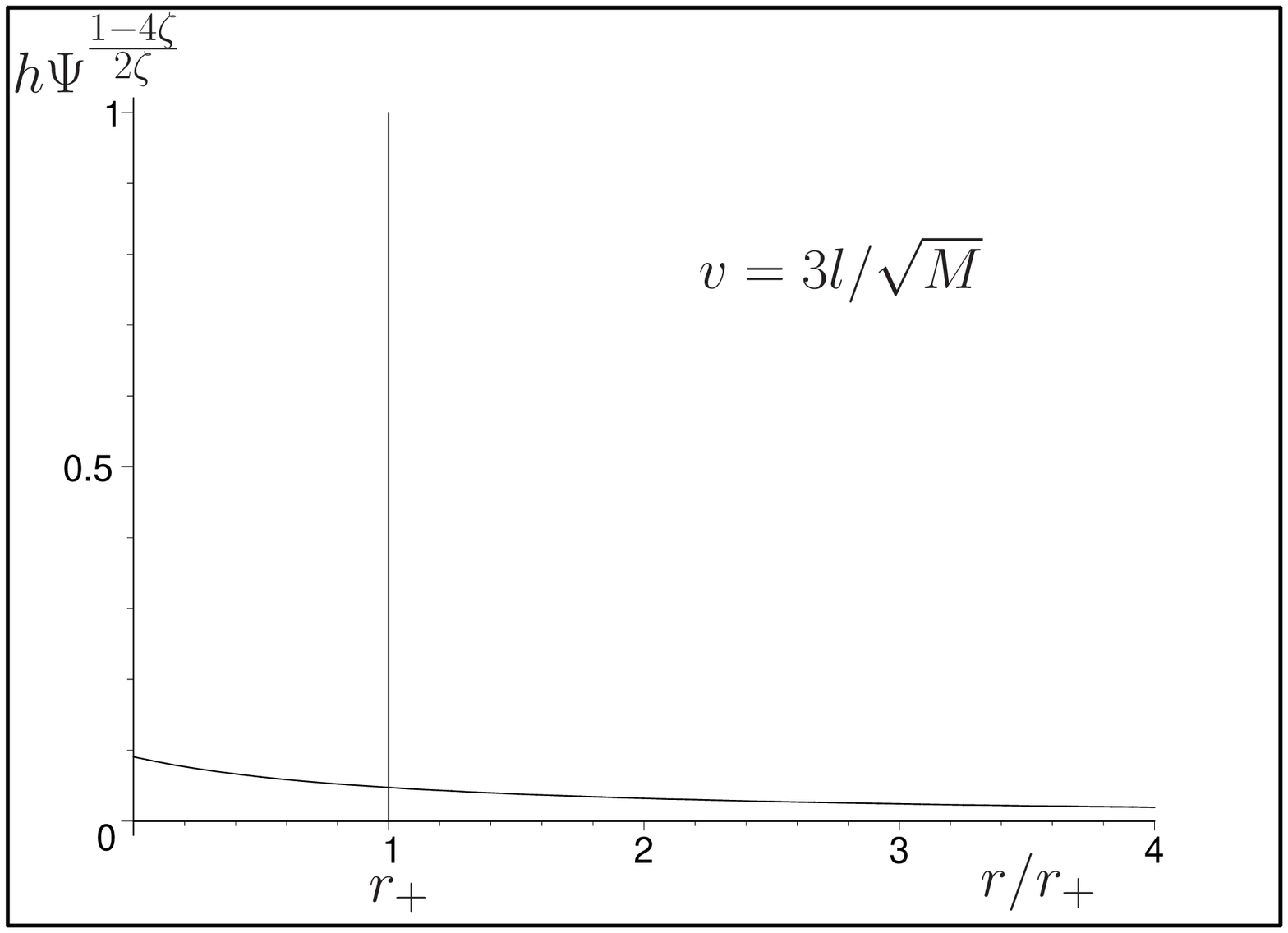}
  \includegraphics[width=5.4cm]{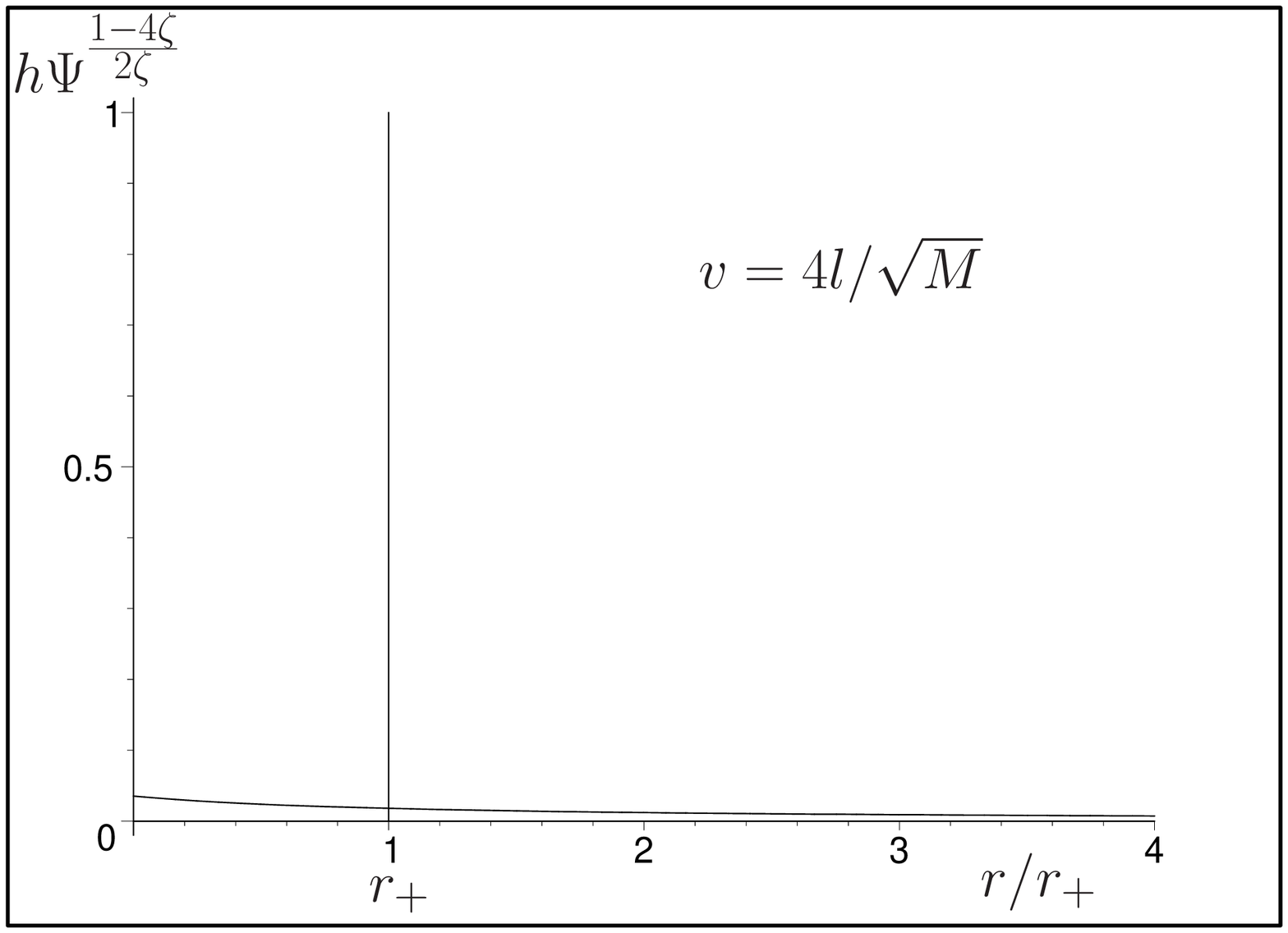}
  \includegraphics[width=5.4cm]{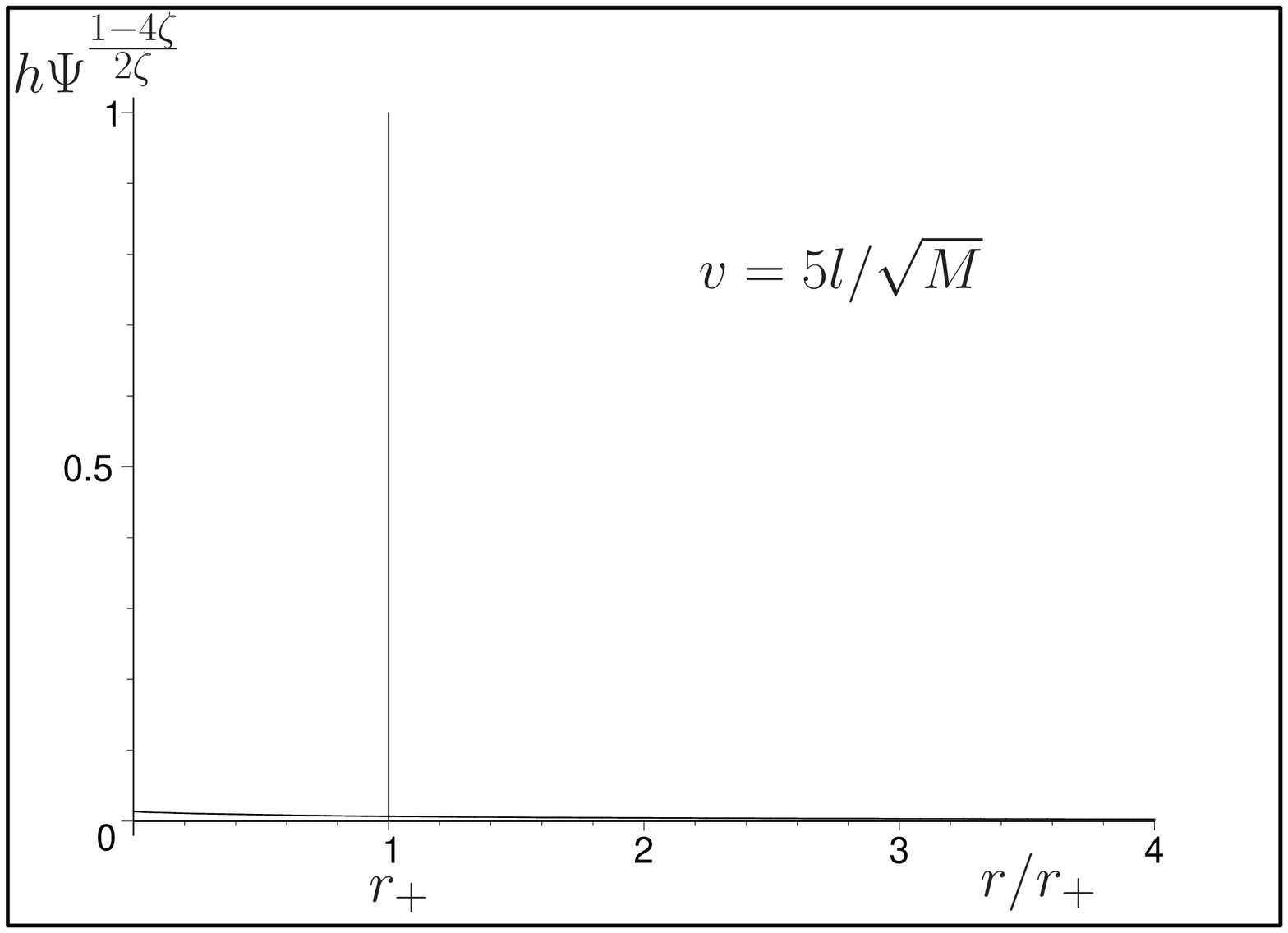}
\caption{\label{fig:Psi(v,r)}Sequence exhibiting the evolution of
the stealth field as time, $v$, increases for nonminimal couplings
in the range $0<\zeta<1/4$. All the graphs were made for
$K=h/\sqrt{M}l$.}
\end{figure}

We would like to stress the novel features of the stealth
configurations in comparison with similar solutions presented
previously (see
\cite{Natsuume:1999cd,Ayon-Beato:2001sb,Henneaux:2002wm,Gegenberg:2003jr}).
First, here a self--interaction potential is included. This
circumvents the problem that for $U(\Psi)=0$, Eq.~(\ref{eq:U}) would
become a constraint severely restricting the parameters of the
problem. In particular, equating to zero the right hand side of
(\ref{eq:U(Psi)}) implies that a nontrivial stealth solution would
only exist if $\zeta=1/8$ or $\zeta=1/6$, and furthermore, only for
$h=0=M$ \cite{Ayon-Beato:2001sb}. Hence, the inclusion of the
self--interaction allows to have stealth solutions for any value of
the nonminimal coupling parameter, and for any black hole mass.

Another important feature of the stealth solutions is the time
dependence since the solutions discussed in
Refs.~\cite{Natsuume:1999cd,Ayon-Beato:2001sb,Henneaux:2002wm,Gegenberg:2003jr}
were stationary and they only exist for the zero mass BTZ geometry.
This stationary solutions belong to a special class, which is
obtained from Eq.~(\ref{eq:feq}) for $f(t)=K=\mathrm{const.}$ and
$M=0$. This is equivalent to taking the limit $M\rightarrow0$ in
Eqs.~(\ref{eq:Psi(t,r)}) and (\ref{eq:Psi(t,r)1/4}). The time
dependence is strictly required in order to have stealth solutions
when $M\neq0$, as can be concluded from Eq.~(\ref{eq:Ptt}). In
relation with this, one can ask if allowing a nontrivial angular
dependence we can also allow for nonzero angular momentum. The
answer to this question, however, is negative. Including an angular
dependence in $\Psi$ makes the system
$T_\mu^{~\nu}(\bm{g}_{\mathrm{BTZ}},\Psi)=0$ although obviously
quite involved, it can be integrated again. The condition that these
solutions respect the identification $\phi=\phi+2\pi$ globally
implies that $\partial_{\phi}\Psi=0$. Hence, expression
(\ref{eq:Psi(t,r)}) is the most general solution on the $2+1$ black
hole.

The stealth solutions presented here have no influence on the
gravitational field, but an important issue is whether their
quantum fluctuations would produce back reaction on the geometry
or not. The question is whether quantum corrections to the black
hole would produce a nonzero expectation value of the quantum
energy--momentum operator.

\begin{acknowledgments}
We are thankful to A.~Garc\'{\i}a, M.~Hassaine, V.~Husain,
D.~Robinson, and R.~Troncoso for many enlightening and helpful
discussions. This work was partially funded by FONDECYT Grants
1040921, 1020629, 1010446, 1010449, 1010450, 1051064, 1051056,
7020629, and 7040190 from, CONACyT Grants 38495E and 34222E,
CONICYT/CONACyT Grant 2001-5-02-159 and Fundaci\'on Andes Grant
D-13775. The generous support of Empresas CMPC to the Centro de
Estudios Cient\'{\i}ficos (CECS) is also acknowledged. CECS is a
Millennium Science Institute and is funded in part by grants from
Fundaci\'on Andes and the Tinker Foundation.
\end{acknowledgments}

\end{document}